\documentclass[letterpaper, 10 pt, conference]{ieeeconf}  % Comment this line out
\IEEEoverridecommandlockouts                              % This command is only
\newcommand{\vf}{\mathbf f}

\newcommand{\vl}{\mathbf l}

\newcommand{\vq}{\mathbf q}
\newcommand{\vr}{\mathbf r}
\newcommand{\vs}{\mathbf s}

\newcommand{\vu}{\mathbf u}

\newcommand{\vw}{\mathbf w}
\newcommand{\vx}{\mathbf x}

\newcommand{\vA}{\mathbf A}
\newcommand{\vB}{\mathbf B}
\newcommand{\vC}{\mathbf C}

\newcommand{\vG}{\mathbf G}

\newcommand{\vL}{\mathbf L}

\newcommand{\vP}{\mathbf P}
\newcommand{\vQ}{\mathbf Q}
\newcommand{\vR}{\mathbf R}
\newcommand{\vS}{\mathbf S}

\usepackage{graphics} % for pdf, bitmapped graphics files
\usepackage{epsfig} % for postscript graphics files
\usepackage{mathptmx} % assumes new font selection scheme installed
\usepackage{times} % assumes new font selection scheme installed
\usepackage{amsmath} % assumes amsmath package installed
\usepackage{amssymb}  % assumes amsmath package installed
\usepackage{algorithm,algorithmic}

\title{\LARGE \bf
Risk Sensitive, Nonlinear Optimal Control: Iterative Linear Exponential-Quadratic Optimal Control with Gaussian Noise }

\author{Farbod Farshidian and Jonas Buchli% <-this % stops a space
\thanks{F. Farshidian and J. Buchli are with the
Agile \& Dexterous Robotics Lab at the Institute of Robotics and Intelligent
Systems, ETH Z\"urich, Switzerland. \{farbodf, buchlij\}@ethz.ch}%
}

\begin{document}

\maketitle
\thispagestyle{empty}
\pagestyle{empty}

%%%%%%%%%%%%%%%%%%%%%%%%%%%%%%%%%%%%%%%%%%%%%%%%%%%%%%%%%%%%%%%%%%%%%%%%%%%%%%%%
\begin{abstract}

  In this contribution, we derive ILEG, an iterative algorithm to find risk
  sensitive solutions to nonlinear, stochastic optimal control problems. The
  algorithm is based on a linear quadratic approximation of an exponential risk
  sensitive nonlinear control problem. ILEG allows to find risk sensitive
  policies and thus generalizes previous algorithms to solve nonlinear optimal
  control based on iterative linear-quadratic methods. Depending on the setting
  of the parameter controlling the risk sensitivity, two different strategies on
  how to cope with the risk emerge. For positive-value parameters, the control
  policy uses high feedback gains whereas for negative-value parameters, it uses a
  robust feedforward control strategy (a robust plan) with low gains. These
  results are illustrated with a simple example.  This note should be considered
  as a preliminary report.

\end{abstract}

%%%%%%%%%%%%%%%%%%%%%%%%%%%%%%%%%%%%%%%%%%%%%%%%%%%%%%%%%%%%%%%%%%%%%%%%%%%%%%%%
\section{Introduction}
The advent of cheap and fast processors and the increasing application of 
complex embedded systems, like robots, has made computational methods for
controller design very appealing. Optimal control theory provides a set of
tools to establish this connection between numerical computation and
control. Among the wealth of numerical methods proposed in the optimal
control framework, the Sequential, Linear, Quadratic (SLQ) algorithms are of a
significant importance because of their computational efficiency. The main idea
behind SLQ is to approximate the original nonlinear optimal control problem by a
series of local Linear-Quadratic (LQ) problems. Based on the solutions
of this local problems we
can iteratively improve the solution to the original nonlinear problem.
Algorithms with this spirit are reported in \cite{mayne66, dunn89, sideris05,todorov05}.

One of the main drawbacks of standard SLQ formulation is that the resulting controller for a
stochastic problem with additive process noise is identical to the one which
is obtained by neglecting noise. In other words, the derived SLQ controller is
independent of the process noise statistics. This is known as the certainty
equivalence principle and it stems from the fact that the cost function
in an LQ problem only considers the mean of the given performance index.

In order to deal with this issue, it is necessary to include higher order
statistics of the performance index into the cost function. However a na\"ive
implementation of this idea only increases the nonlinearity of the problem.
One interesting approach to incorporate the higher order statistics is proposed
by Jacobson \cite{jacobson73}. In his risk sensitive control scheme which uses the
expectation of the exponential transformation of the performance index, he showed
that the optimal controller is sensitive to the noise statistics. More
importantly the computational difficulty of calculating this risk sensitive
controller is the same as the original LQ problem for the expectation of the
performance index.

In this paper, we will use the SLQ idea to sequentially approximate the nonlinear
problem with local LQ subproblems. However, instead of the conventional approach,
we will use the risk sensitive method to design the local controllers for the LQ
subproblems. Therefore, the proposed algorithm in this paper will iteratively
approximate the nonlinear problem with a risk sensitive LQ problem. The rest of
this paper is organized as follows; First, we show the relationship between the
solution of the risk sensitive optimal control problem to the one with the
conventional cost function. Then we will derive the theory behind our algorithm.
Finally, we will illustrate its performance on a continuous cliff world problem.

%%%%%%%%%%%%%%%%%%%%%%%%%%%%%%%%%%%%%%%%%%%%%%%%%%%%%%%%%%%%%%%%%%%%%%%%%%%%%%%%
\section{Motivation}
Consider the following general stochastic nonlinear optimal control problem.
\begin{equation} \label{eq:wiener_process}
d\vx_t = \left( \vf(t,\vx_t) + \vG(t,\vx_t)\vu_t \right)dt + \vC(t,\vx_t) d\vw_t \quad \vx(0)=\vx_0
\end{equation}
where $d\vw_t$ is a Brownian motion with zero mean and covariance $\Sigma dt$ and
the cost function is defined as
\begin{equation} \label{eq:general_cost_funtion}
J = \min_{\pi} \mathbb{E} \left\{ \mathcal{J(\pi)} \right\}
\end{equation}
$\mathcal{J(\pi)}$ is the performance index which is in general a random variable
and a functional of the control policy, $\pi$. $\mathbb{E}$ represents the
expectation with respect to this random variable.

This general optimal control problem does not have an analytical
solution, except for a few special cases. One of these cases is a linear system
with a quadratic cost function. However, as uncertainty equivalence principle
states the solution to this LQ problem does not consider the
stochastic characteristic of the problem, i.e. the designed control policy is
indifferent to the stochasticity of the problem. The reason is that the LQ
problem only considers the mean of the cost and ignores the higher order
momenta. A possible solution could be to add a measure of variance to the
regular cost function. Unfortunately, the resulting problem is not anymore an LQ
problem and there is no efficient algorithm to find the solution.

Following the idea of incorporating higher order momenta of the cost
function, we can consider the following family of exponential
functions:
\begin{equation} \label{eq:general_exp_cost_funtion}
J = \min_{\pi} \mathbb{E} \left\{ \exp \left[ \sigma \mathcal{J(\pi)} \right] \right\}
\end{equation} 
where $\sigma$ is a real valued parameter. 

\textbf{\textit{Corollary 1}}: The logarithm of the cost function in Equation \eqref{eq:general_exp_cost_funtion} can be expanded as
\begin{equation}
\frac{1}{\sigma} \log [J] = \mathbb{E}[\mathcal{J}^*] + \frac{\sigma}{2} \mathbb{\mu}_2[\mathcal{J}^*] + \frac{\sigma^2}{6} \mathbb{\mu}_3[\mathcal{J}^*] + ...
\end{equation}
where $\mathbb{\mu}_2$ and $\mathbb{\mu}_3$ are the variance and the skewness of $\mathcal{J^*}$ (the cost of the optimal policy) respectively. \\
\textit{Proof}: see Appendix A.

Corollary 1 shows that by using the exponential cost function family, we can
incorporate the momenta of higher orders of the original cost function
momentum in the optimal control problem. Fortunately, like for the LQ problem, we can find an
analytical solution for the optimal control problem with linear dynamics and a
cost function defined as the exponential of a quadratic cost. In this paper we
will investigate this class of optimal control problems in more detail. We will
call this problem the \emph{Linear (linear dynamics), Exponential-quadratic
(Exponential-quadratic cost) problem with Gaussian process noise} or in short
``LEG'' optimal control problem.

In the next section, we will devise a dynamic programming approach to find the
optimal controller for the general problem with exponential cost. Furthermore, we
will show that this family of problems includes the common (i.e. with respect to
the mean) optimal control problem as a special case for a specific choice of a
parameter.

%%%%%%%%%%%%%%%%%%%%%%%%%%%%%%%%%%%%%%%%%%%%%%%%%%%%%%%%%%%%%%%%%%%%%%%%%%%%%%%%
\section{Problem formulation}
First, assume a general optimal control problem with the following exponential
cost function:
\begin{equation} \label{eq:cost_funtion}
J = \min\limits_{ \vu_0 \to \vu_{t_f}} \mathbb{E}\left\{ \exp\left[ \sigma \left(\Phi_f(\vx_{t_f})+ \int_0^{t_f} { L(t,\vx_t,\vu_t)dt}\right) \right] \right\}
\end{equation}
where $L(t,\vx_t,\vu_t)$ is defined as
\begin{equation} \label{eq:cost_funtion_L}
L(t,\vx,\vu) = \Phi(t,\vx) + \frac{1}{2} \vu^T \vR(t,\vx) \vu + \vu^T \vr(t,\vx)
\end{equation}
$\Phi(t,\vx)$ is a general nonlinear function and the state trajectories are
generated through the stochastic system defined by Equation \eqref{eq:wiener_process}.

\textbf{\textit{Theorem 1}}: The solution to the optimal control problem defined
in Equations \eqref{eq:wiener_process} and \eqref{eq:cost_funtion} is
\begin{align}
& J = \exp \left[ \sigma \Psi(0,\vx_0) \right] \\
&\vu^*(t,\vx) =  \vR(t,\vx)^{-1} \left( \vr(t,x) + \vG^T(t,\vx) \nabla_{\vx}\Psi(t,\vx) \right)
\label{eq:optimal_control}
\end{align}
where $\Psi(t,\vx)$ is the solution to the following partial differential equation (PDE)
\begin{align} \label{eq:nonlinear_utility_pde}
- \partial_t\Psi &= \Phi -\frac{1}{2}\vr^T \vR^{-1} \vr + \nabla_{\vx}\Psi^T \left( \vf - \vG\vR^{-1}\vr \right) - \notag \\
&  \frac{1}{2} \nabla_{\vx}\Psi^T \big(  GR^{-1}G^T - \sigma \vC\Sigma\vC^T \big) \nabla_{\vx}\Psi + \frac{1}{2} Tr [\nabla_{\vx\vx}\Psi \vC\Sigma\vC^T]
\end{align}
with boundary condition $\Psi(t_f,\vx) = \Phi_f(x)$ (in the interest
of compact notation, we dropped the functionality with respect to $t$ and $\vx$). \\
\textit{Proof}: see Appendix B.

We call the PDE in Equation \eqref{eq:nonlinear_utility_pde} the \emph{extended
Hamilton-Jacobi-Bellman Equation} or in short \emph{extended HJB Equation}.
This equation forms the basis for deriving our algorithm which iteratively
approximates a general nonlinear exponential optimal control problem by LEG
optimal control  in order to approximate the solution in an efficient manner.
Before continuing to the next section, we will take a look at the relationship
between the exponential and the common optimal control problem.

\textbf{\textit{Note}}: If $\sigma$ approaches zero, the optimal control
policy in Equation \eqref{eq:optimal_control} and the value function in Equation
\eqref{eq:nonlinear_utility_pde} are the solution to the common optimal control
problem with the following cost function.
\begin{equation}
J = \min\limits_{ \vu_0 \to t_f} \mathbb{E}\left\{  \Phi_f(\vx_{t_f})+ \int_0^{t_f} { L(t,\vx_t,\vu_t)dt} \right\}
\end{equation}
\textit{Proof}: This can be easily verified by putting $\sigma=0$ and comparing
it with the common HJB equation.

This shows that the exponential optimal control problem converts to the regular
optimal control problem for $\sigma$ equal to zero.

%%%%%%%%%%%%%%%%%%%%%%%%%%%%%%%%%%%%%%%%%%%%%%%%%%%%%%%%%%%%%%%%%%%%%%%%%%%%%%%%
\section{Iterative Linear Exponential-quadratic Optimal Control under Gaussian Process Noise: ILeg} \label{sec:continuous_ILEQG}
ILEG (\textbf{I}terative, \textbf{L}inear, \textbf{E}xponential-quadratic optimal
control under \textbf{G}aussian process noise) is an iterative optimization
method for solving the optimal control problem for a general nonlinear system
with an exponential cost function which is affected by Gaussian process noise.
ILEG designs locally-optimal feedback control for nonlinear, stochastic,
continuous-time systems. Given an initial, feasible sequence of control inputs,
we iteratively obtain a local linear approximation of the system dynamics and a
exponential-quadratic approximation of the cost function, and then incrementally
improve the control law, until we converge to a local minimum. In that sense it
is closely related to previous approaches to solve nonlinear optimal control
algorithms with iterative LQ methods \cite{sideris05,todorov05} with the key
difference that ILEG is risk sensitive and generalizes previous algorithms.

Lets assume we are in iteration $n$ of the algorithm and $\vx^n(t)$ and
$\vu^n(t)$ are respectively the state and control input trajectories generated
through implementation the latest optimized controller. Then we approximate
system dynamics with a time varying linear system along these trajectories and
the cost function with the exponential quadratic function as follows
\begin{align} \label{eq:dynamics_linear_approximation}
& d(\delta\vx_t) = \left( \vA_t\delta\vx_t + \vB_t\delta\vu_t \right)dt + \vC_t d\vw_t \\
& \vA_t = \frac{\partial \vf(t,\vx^n(t))}{\partial \vx(t)}  + \frac{\partial \vG(t,\vx^n(t))}{\partial \vx(t)}\vu^n(t) \label{eq:A} \\
& \vB_t = \vG(t,\vx^n(t)) \label{eq:B}  \\
& \vC_t = \vC(t,\vx^n(t)) \label{eq:C} 
\end{align}
and the quadratic approximation of the cost function
\begin{equation} \label{eq:cost_quadratic_approximation}
J \approxeq \min\limits_{ \vu_{0 \to t_f}} \mathbb{E}\left\{ \exp\left[ \sigma \left(\tilde{\Phi}_f(\vx_{t_f})+ \int_0^{t_f} { \tilde{L}(t,\vx_t,\vu_t)dt}\right) \right] \right\}
\end{equation}
with
\begin{align}
&\tilde{\Phi}_f(\vx) = \ q_f + \vq_f^T \delta\vx + \frac{1}{2} \delta\vx^T \vQ_f \delta\vx \label{eq:cost_quadratic_approximation_detail_1} \\
& \tilde{L}(t,\vx,\vu) = q_{t} + \vq_{t}^T \delta\vx_{t} + \vr_{t}^T \delta\vu_{t} + \frac{1}{2} \delta\vx_{t}^T \vQ_{t} \delta\vx_{t} + \delta\vx_{t}^T \vP_{t} \delta\vu_{t}  \notag \\ 
& \hspace{15mm} + \frac{1}{2} \delta\vu_{t}^T \vR_{t} \delta\vu_{t} \label{eq:cost_quadratic_approximation_detail_2}
\end{align} 
where $\delta \vx_t = \vx(t)-\vx^n(t)$ and $\delta\vu_t = \vu(t)-\vu^n(t)$ and $q_t$, $\vq_t$, $\vr_t$, $\vQ_t$, $\vP_t$, and $\vR_t$ are the coefficients of the Taylor expansion of the cost function over the nominal trajectory.

\textbf{\textit{Theorem 2}}: The solution to the optimal control problem defined in Equations\! (\ref{eq:dynamics_linear_approximation}-\ref{eq:cost_quadratic_approximation_detail_2}) exists if $\left( \vB_t\vR_t^{-1}\vB_t^T -\sigma \vC_t\Sigma\vC_t^T \right)$ is positive semidefinite for all \textit{t} and the solution can be found as follows
\begin{align}
-\dot{\vS}_t =& \vQ_{t} + \vA_t^T \vS_t + \vS_t^T \vA_t - \left( \vP_t^T + \vB_t^T \vS_t \right)^T \vR_t^{-1} \left( \vP_t^T + \vB_t^T \vS_t \right) \notag \\
 &+ \sigma \vS_t^T \vC_t\Sigma\vC_t^T \vS_t  \label{eq:riccati_Sm}\\
-\dot{\vs}_t =& \vq_{t} + \vA_t^T \vs_t - \left( \vP_t^T + \vB_t^T \vS_t \right)^T \vR_t^{-1} \left( \vr_t + \vB_t^T \vs_t \right) \notag \\ 
&+ \sigma  \vS_t^T \vC_t\Sigma\vC_t^T \vs_t \label{eq:riccati_Sv}\\
-\dot{s}_t =& q_{t} - \frac{1}{2} \left( \vr_t + \vB_t^T \vs_t \right)^T \vR_t^{-1} \left( \vr_t + \vB_t^T \vs_t \right) + \frac{1}{2} Tr\left[ \vS_t \vC_t\Sigma\vC_t^T \right] \notag \\
&+ \frac{\sigma}{2} \vs_t^T \vC_t\Sigma\vC_t^T \vs_t   \label{eq:riccati_s}
\end{align}
with the final values $\vS_{t_f} = \vQ_{f}$, $\vs_{t_f} = \vq_{f}$, and $s_{t_f} = q_{f}$. The optimal control is
\begin{align}
&\delta\vu^*(t,\vx) = \vl(t) + \vL(t) \delta\vx \label{eq:optimal_control_update}\\
&\vl(t) = - \vR_t^{-1} \left( \vr_t + \vB_t^T \vs_t \right) \label{eq:optimal_control_l}\\
&\vL(t) = - \vR_t^{-1} \left( \vP_t^T + \vB_t^T \vS_t \right) \label{eq:optimal_control_L}
\end{align}
\textit{Proof}: see Appendix C.

%%%%%%%%%%%%%%%%%%%%%%%%%%%%%%%%%%%%%%%%%%%%%%%%%%%%%%%%%%%%%%%%%%%%%%%%%%%%%%%%
\section{Summary of the ILEG Algorithm}
Algorithm \ref{alg:ileg} summarizes the ILEG algorithm described in the previous
section. This algorithm assumes the system dynamics and the exponential cost
function as given. It also requires to define a parameter named $\sigma$. As we
stated in the Theorem 2, the matrix expression $\left( \vB_t\vR_t^{-1}\vB_t^T
-\sigma \vC_t\Sigma\vC_t^T \right)$ should be always positive semi-define which
imposes an upper bound over $\sigma$. In the next section we will discuss the
effect of this parameter in more details.

In each iteration of this algorithm, we need to forward integrate the noise-free
system dynamics using the latest update of the controller. Then we approximate
the system dynamics and the cost function along the forward-integrated trajectories. The
algorithm use a linear approximation for the system dynamics and an
exponential-quadratic approximation for the cost function.

In the next step, we solve the approximated LEG problem using the results
from Theorem 2. This solution gives us an update to the optimal control policy.
Finally we should iterate this process until a termination condition is fulfilled.
\begin{algorithm}[tpb]
\caption{ILEG Algorithm}
\label{alg:ileg}
\begin{algorithmic}
\scriptsize 
	\STATE \textbf{Given}
	\STATE - System dynamics in Equation \eqref{eq:wiener_process} 
	\STATE - Cost function in Equation \eqref{eq:cost_funtion}
	\STATE - Choose $\sigma$ in the allowed range
	\STATE \textbf{Initialization}
	\STATE - Initialize the controller with a stable control law, $\mathbf{\pi}(t,\vx)$
	\REPEAT
	\STATE - Forward integrate the system dynamics:
	\STATE $\tau: \vx^n(0),\vu^n(0),\vx^n(1),\vu^n(1),\dots,\vx^n(t_f-1),\vu^n(t_f-1),\vx^n(t_f)$
	\STATE - Compute the linear approximation of the system dynamics along the nominal trajectory $\tau$, Equations\! (\ref{eq:A}-\ref{eq:C})
	\STATE - Compute the quadratic approximation of the cost function along the nominal trajectory $\tau$, Equations\! (\ref{eq:cost_quadratic_approximation_detail_1}-\ref{eq:cost_quadratic_approximation_detail_2}) 
	\STATE - Solve the final value differential Equations\! (\ref{eq:riccati_Sm}-\ref{eq:riccati_s})
	\STATE - Update the control law: $\mathbf{\pi}(t,\vx) = \vu^n(t) + \vl(t) + \vL(t) \left(\vx(t)-\vx^n(t)\right)$
	\UNTIL{a termination condition is matched}
\end{algorithmic}
\end{algorithm} 

%%%%%%%%%%%%%%%%%%%%%%%%%%%%%%%%%%%%%%%%%%%%%%%%%%%%%%%%%%%%%%%%%%%%%%%%%%%%%%%%
\section{Numerical Example}
\begin{figure} [bpb]
\centering
\includegraphics[width=0.5\textwidth]{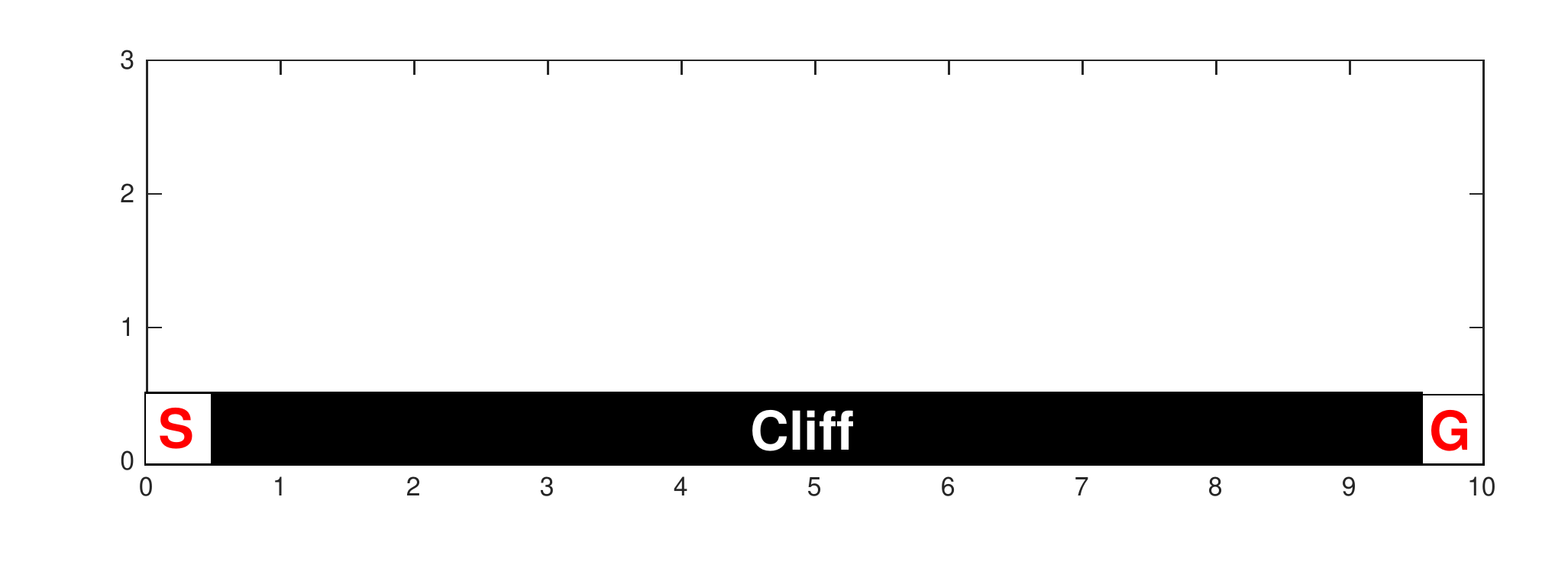}
\caption{A continuous cliff world. S and G indicate the start and goal
  position respectively. Moving through the white region induces low
  cost, while ``falling'' over the cliff induces very high cost.}
\label{fig:cliff_world}
\end{figure}

In this section, we will show some preliminary results of the ILEG
implementation on a continuous cliff world problem. In this problem a point mass
(1kg) should be navigated from one corner of a rectangle area to the other while
at the border of the area there is a cliff (Figure \ref{fig:cliff_world}). The
mass point motion is influenced by a Brownian motion on both the X and the Y
directions. However the noise standard deviation (SD) in the Y directions is 10 times higher which increases the chances of falling off the cliff. The goal of
this problem is to design a controller which can navigate the point mass form the
start point to the goal point with minimum control effort without falling.

In order to formulate this problem as an optimal control problem as defined by 
Equations \eqref{eq:wiener_process} and \eqref{eq:cost_funtion}, we should
replace the hard constraint of the cliff by a soft constraint which penalizes the
distance of the mass from the cliff. Thus, we define the following cost function
for this problem
\begin{align}
\label{eq:phi_cliff_world} 
& \Phi_f(\vx) = 100(x-10)^2 + 100y^2 + 10(\dot{x}^2 + \dot{y}^2) \\
\label{eq:l_cliff_world} 
& L(t,\vx,\vu) = \frac{0.1}{(0.1y+1)^{10}} + u_x^2 + 0.01u_y^2
\end{align} 
Equation \eqref{eq:phi_cliff_world} is the terminal cost at $t_f=3$ which puts a
high penalty for deviating from the goal state, $[10,0]^T$, at time 3[sec]. It
also penalizes the point mass speed at the final time. Therefore, the final cost
encourages the point mass to reach and stop at the goal state within 3 seconds.
In Equation \eqref{eq:l_cliff_world}, the first term is a penalty term for
falling off the cliff. Finally, the last two terms add cost for the exerted
control forces in each motion directions. Notice that since the noise in Y
direction has higher standard deviation, we penalize the controller less for the
effort to confront the noise.

Although the point mass in this problem has linear system dynamics, the defined
cost function is nonlinear. Consequently the optimal control problem defined by
this cost function is nonlinear. We use ILEG on this problem to find the optimal
policy. The algorithm converges after few iterations. The resulting control
policy shows different characteristics depending on the chosen parameter 
$\sigma$. In general, $\sigma$ has an upper limit beyond which the designed
policy will be unstable. In this cliff world problem, this limit is 50. Here, we
implemented the ILEG algorithm for 5 different choices of $\sigma$.

\begin{figure} [tpb]
\centering
\includegraphics[width=0.5\textwidth]{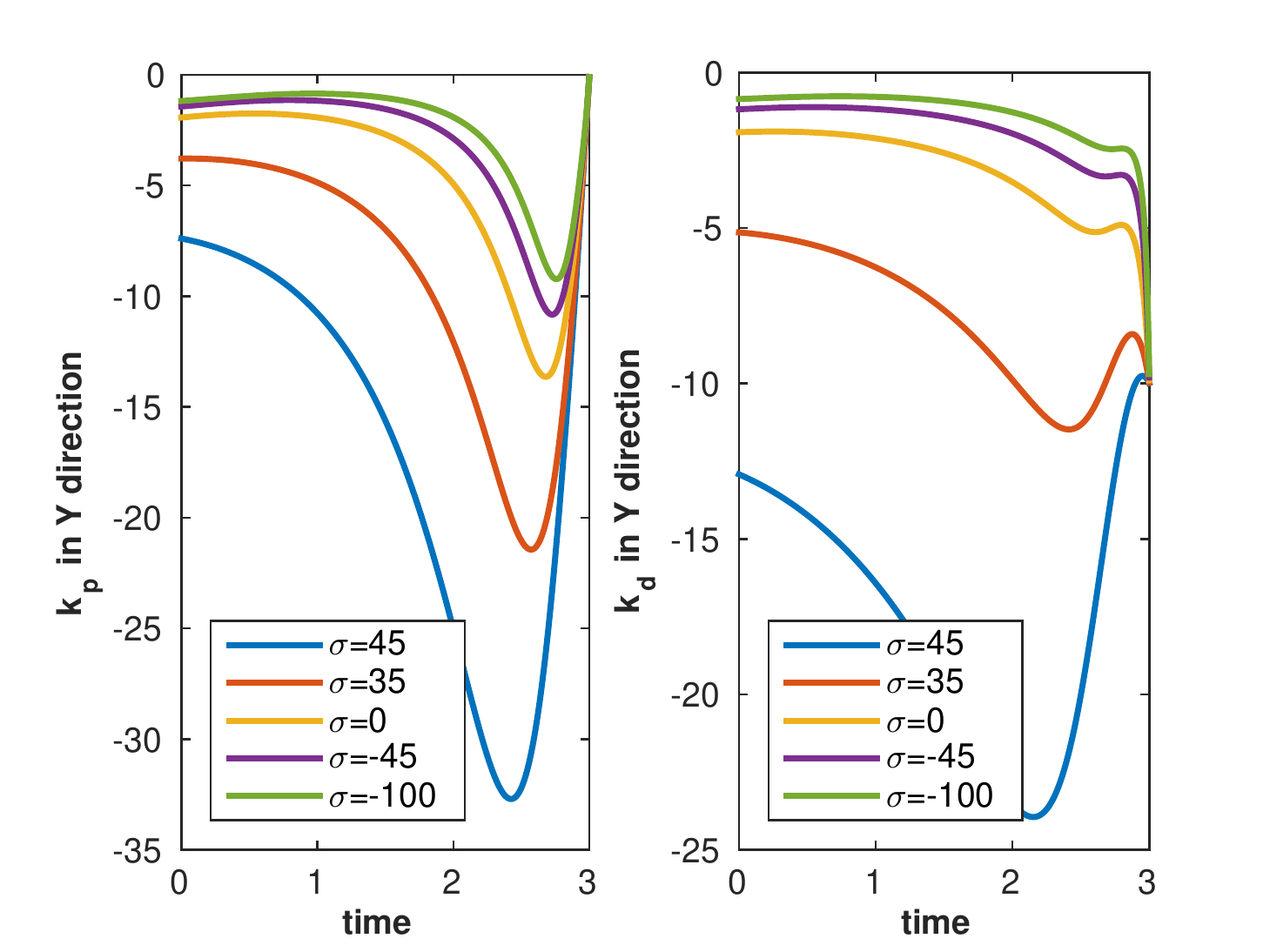}
\caption{Y direction controller gains for 5 different $\sigma$ values, namely 45, 35, 0, -45, and -100. The left plot illustrates the changes of the proportional gains in the course of time and the right plot shows the derivative gains.}
\label{fig:gains_y_direction}
\end{figure}

\begin{figure*} [tpb]
\centering
\includegraphics[width=0.7\textwidth]{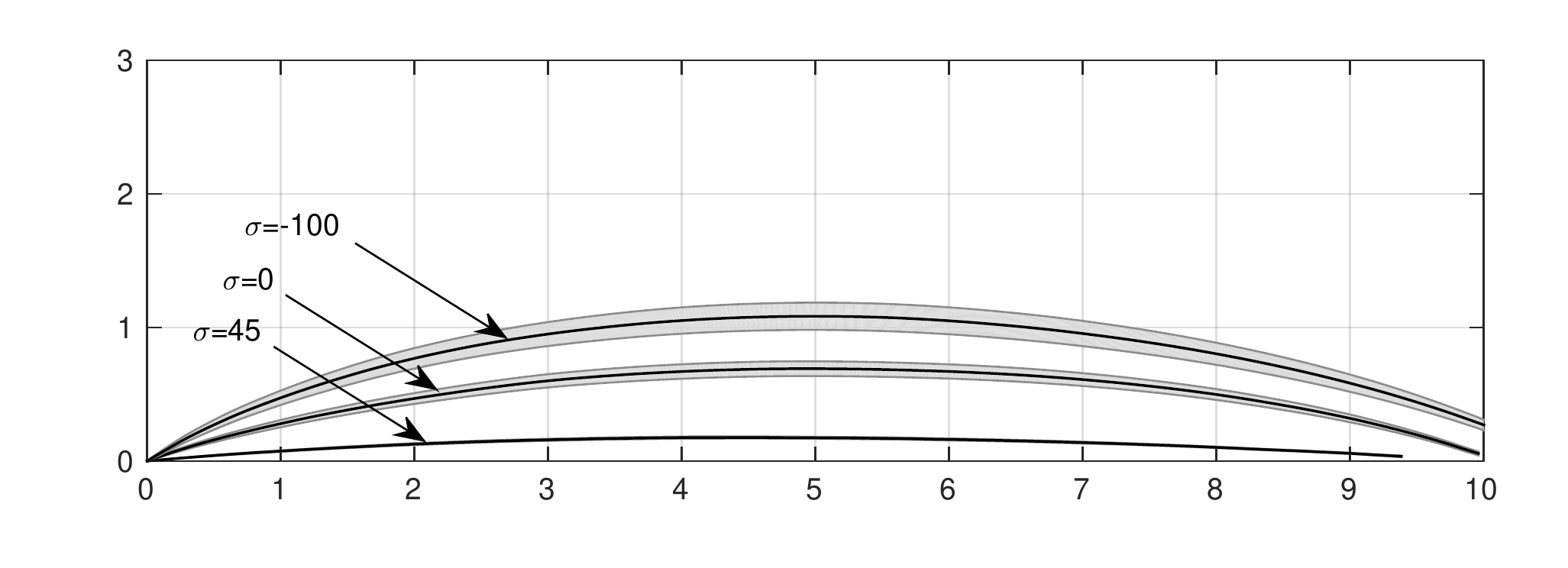}
\caption{The traversed path of the point mass using controllers with 3 different $\sigma$ values, namely 45, 0, and -100. The shaded area is 15 percent SD of the trajectories}
\label{fig:sigma_comparison}
\end{figure*}

Figure \ref{fig:gains_y_direction} demonstrates the changes of the feedback gains
over time in the Y direction. As expected, by decreasing $\sigma$ from
$\sigma=+45$ to $\sigma=-100$ the absolute value of the gains decreases
monotonically. For $\sigma=0$, the value of $\sigma$ which the controller does
not take the stochasticity of the problem into account (it is the solution to the
non exponential cost function). As Figure \ref{fig:gains_y_direction} shows by
increasing $\sigma$ to positive values the controller uses higher gains to reduce
the variance of the generated motions. However, if we decrease $\sigma$ to
negative values, the controller uses lower gains and therefore the motion
generated under this controller will have higher variations. In order to
compensate for these higher variations which can cause the point mass to fall off
the cliff, whereas the $\sigma$-negative controller will choose a more
conservative path. In other words, to deal with the uncertainty, the controller
prefers a safer plan over a stiffer controller. Figure \ref{fig:sigma_comparison}
illustrates this for three different $\sigma$ values.

As in Figure \ref{fig:sigma_comparison}, the positive $\sigma$ takes a shorter
path than the negative one while the negative $\sigma$ chooses a safer path. In
this figure the shaded error--bands are a measure of the path variations under the
system noise. We see that, since the negative $\sigma$ has lower gains, it has a
wider error--band than the positive one.

%%%%%%%%%%%%%%%%%%%%%%%%%%%%%%%%%%%%%%%%%%%%%%%%%%%%%%%%%%%%%%%%%%%%%%%%%%%%%%%%
\section{CONCLUSIONS AND FUTURE WORK}
In this preliminary work, we have introduced an iterative optimal control
algorithm named as ILEG. ILEG iteratively approximates the system dynamics and
the cost function by a linear system and an exponential-quadratic cost
respectively. Then it efficiently solves the approximated LEG subproblems. We
showed that the advantages of using exponential cost function instead of a
regular one is that the higher order momenta of the performance index are also
considered during the optimization.
  
An interesting aspect of the ILEG algorithm is that it introduces an algorithmic
parameter which can control the behavior of the optimal control. By setting this
parameter to zero, ILEG basically reduces to the well-known SLQ. However by
setting this parameter to a positive or a negative value, we can obtain two
different types of policies. For the positive-value parameter the control policy
mostly relies on the error feedback signal, using high gains ('stiff controls')
while in the negative-value parameter the control policy contains a robust plan
(forward controls), using lower gains.

\subsection{Future Work}
This work is currently in its early stage. The effect of the $\sigma$ parameter
should be studied through more analytical methods rather than a numerical
example. Furthermore, even though Algorithm \ref{alg:ileg} imposes an upper bound
on $\sigma$, it is not totally clear that this is the only restriction over
$\sigma$. Questions like the stability of the designed controller under different
values of $\sigma$ should be also addressed. Last but not least, the proposed
algorithm should be implemented on more practical examples.

%%%%%%%%%%%%%%%%%%%%%%%%%%%%%%%%%%%%%%%%%%%%%%%%%%%%%%%%%%%%%%%%%%%%%%%%%%%%%%%%
\section{ACKNOWLEDGMENTS}
This research has been supported in part by a Swiss National Science
Foundation Professorship Award to Jonas Buchli, the NCCR Robotics and a
Max-Planck ETH Center for Learning Systems Ph.D. fellowship to Farbod Farshidian.

%%%%%%%%%%%%%%%%%%%%%%%%%%%%%%%%%%%%%%%%%%%%%%%%%%%%%%%%%%%%%%%%%%%%%%%%%%%%%%%%

%===========================================
% Bibliography
%===========================================
\bibliographystyle{bibliography/IEEEtran}
\bibliography{bibliography/IEEEabrv,bibliography/references}

%%%%%%%%%%%%%%%%%%%%%%%%%%%%%%%%%%%%%%%%%%%%%%%%%%%%%%%%%%%%%%%%%%%%%%%%%%%%%%%%

\onecolumn

%%%%%%%%%%%%%%%%%%%%%%%%%%%%%%%%%%%%%%%%%%%%%%%%%%%%%%%%%%%%%%%%%%%%%%%%%%%%%%%%

\section{Appendix A}
\textbf{\textit{Corollary 1}}: The cost function in Equation \eqref{eq:general_exp_cost_funtion} can be expanded as
\begin{equation}
\frac{1}{\sigma} \log [J] = \mathbb{E}[\mathcal{J}^*] + \frac{\sigma}{2} \mathbb{\mu}_2[\mathcal{J}^*] + \frac{\sigma^2}{6} \mathbb{\mu}_3[\mathcal{J}^*] + ...
\end{equation}
where $\mathbb{\mu}_2$ and $\mathbb{\mu}_3$ are the variance and the skewness of
$\mathcal{J^*}$ respectively.

\textbf{\textit{Proof}}: Assuming that the cost associated with one execution of
the optimal policy is $\mathcal{J}^*$. We can show $J$ as
\begin{equation}
J = \mathbb{E}\left\{\exp[\sigma \mathcal{J}^*]\right\}
\end{equation}
Therefore we will have
\begin{equation}
\frac{1}{\sigma} \log [J] = \frac{1}{\sigma} \log \mathbb{E}\left\{\exp[\sigma \mathcal{J}^*]\right\}
\end{equation}
$\log \mathbb{E}\left\{\exp[\sigma \mathcal{J}^*]\right\}$ is the cumulant
generating function of the random variable $\mathcal{J}^*$. By writing the Taylor
series expansion of $\log [J]$, we will have
\begin{equation}
\frac{1}{\sigma} \log [J] = = \sum_{i=1}^{\infty}{\kappa_i \frac{\sigma^{i-1}}{i!}}
\end{equation}
where $\kappa_i$ is the \textit{i}th cumulant of $\mathcal{J}$. Using the fact
that the first three cumulants are mean, variance, and skewness will conclude the
proof.

%%%%%%%%%%%%%%%%%%%%%%%%%%%%%%%%%%%%%%%%%%%%%%%%%%%%%%%%%%%%%%%%%%%%%%%%%%%%%%%%
\section{Appendix B} 
\textbf{\textit{Theorem 1}}: The solution to the optimal control problem defined
in Equations \eqref{eq:wiener_process} and \eqref{eq:cost_funtion} is
\begin{align}
& J = \exp \left[ \sigma \Psi(0,\vx_0) \right] \\
&\vu^*(t,\vx) =  \vR(t,\vx)^{-1} \left( \vr(t,x) + \vG^T(t,\vx) \nabla_{\vx}\Psi(t,\vx) \right)
\label{app_eq:optimal_control}
\end{align}
where $\Psi(t,\vx)$ is the solution to the following partial differential equation (PDE)
\begin{equation} \label{app_eq:nonlinear_utility_pde}
-\partial_t\Psi = \Phi -\frac{1}{2}\vr^T \vR^{-1} \vr + \nabla_{\vx}\Psi^T \left( \vf - \vG\vR^{-1}\vr \right) - \frac{1}{2} \nabla_{\vx}\Psi^T \big( GR^{-1}G^T - \sigma \vC\Sigma\vC^T \big) \nabla_{\vx}\Psi + \frac{1}{2} Tr [\nabla_{\vx\vx}\Psi \vC\Sigma\vC^T]
\end{equation}
with boundary condition $\Psi(t_f,\vx) = \Phi_f(x)$ (to make the equation
shorter, we have dropped the functionality with respect to $t$ and $\vx$).

\textbf{\textit{{Proof}}}: 
In order to solve this optimal control problem, we chose a dynamic programming
approach. First consider a discrete time problem with the system dynamics
described by Equation \eqref{app_eq:discrete_dynamics}
\begin{equation} \label{app_eq:discrete_dynamics}
\vx_{n+1}= \vf(n,\vx_n) + \vG(n,\vx_n)\vu_n  + \vC(n,\vx_n) \vw_n
\end{equation}
where $\vw_n$ is a Gaussian random process with zero mean and covariance
$\mathbb{C}ov(\vw_n,\vw_m) = \Sigma \delta_{n,m}$. The discrete cost function is
also defined as the following
\begin{equation} \label{app_eq:discrete_cost}
J = \min\limits_{ \vu_0 \dots \vu_{N-1}} \mathbb{E}\left\{ \exp\left[ \sigma \left(\Phi(x_N)+ \sum_{n=0}^{N-1} { L(n,\vx_n,\vu_n)}\right) \right] \right\}
\end{equation}
It can be easily shown that the solution to this discrete optimal control problem
can be obtained using Equation \eqref{app_eq:extended_bellman}.
\begin{equation} \label{app_eq:extended_bellman}
V(n,\vx) = \min\limits_{\vu_n} \Big\{\exp \left( \sigma L(t,\vx,\vu_n) \right) \mathbb{E} \left[V(n+1,\vx_{n+1})\right] \Big\},  \qquad V(N,\vx) = \Phi(\vx)
\end{equation}
This equation is called the extended Bellman equation. 

In order to find the solution of the continuous time optimal control problem
defined in Equations \eqref{eq:wiener_process} and \eqref{eq:cost_funtion}, we
should find the equivalent dynamic programming formula. This can be achieved by
discretizing the continuous time equation and using the extended Bellman equation
to find the optimality equation. Then by the use of the Ito lemma, we can derive the
following optimality equation called extended HJB equation.
\begin{equation} \label{app_eq:extended_hjb}
-\partial_t V(t,\vx) = \min\limits_{\vu_t} \left\{ \sigma V(t,\vx) L(t,\vx,\vu_t) + 
\nabla_{\vx}V^{T}(t,\vx) \left( \vf(t,\vx)+\vG(t,\vx)\vu_t \right) + \frac{1}{2} Tr [\nabla_{\vx\vx}V(t,\vx) \vC(t,\vx_t) \Sigma \vC^T(t,\vx_t)] \right\}, \hspace{1mm} V(t_f,\vx) = \Phi(\vx) 
\end{equation}
Using the exponential transformation $V(t,\vx) = \exp(\sigma \Psi(t,\vx))$ in \eqref{app_eq:extended_hjb} we get
\begin{align}
& V(t,\vx) = \exp(\sigma \Psi(t,\vx)) \\
&\partial_t V(t,\vx) = \sigma V(t,\vx) \partial_t \Psi(t,\vx) \\
&\nabla_{\vx}V(t,\vx) = \sigma V(t,\vx) \nabla_{\vx}\Psi(t,\vx) \\
&\nabla_{\vx\vx}V(t,\vx) =  V(t,\vx) \left( \sigma^2 \nabla_{\vx}\Psi(t,\vx)\nabla_{\vx}\Psi^T(t,\vx) + \sigma \nabla_{\vx\vx}\Psi(t,\vx) \right)
\end{align}
Substituting these equations in the extended HJB equation (for the simplicity we
will drop all of the subscripts)
\begin{align}
-\sigma V(t,\vx) \partial_t \Psi(t,\vx) =& \min\limits_{\vu_t} \Big\{ \sigma V(t,\vx) L(t,\vx,\vu_t) + 
\sigma V(t,\vx) \nabla_{\vx}\Psi^T(t,\vx) \left( \vf(t,\vx)+\vG(t,\vx)\vu_t \right) + \notag \\ 
&\frac{1}{2} Tr [ V(t,\vx) \left( \sigma^2 \nabla_{\vx}\Psi(t,\vx)\nabla_{\vx}\Psi^T(t,\vx) + \sigma \nabla_{\vx\vx}\Psi(t,\vx) \right) \vC(t,\vx_t) \Sigma \vC^T(t,\vx_t)] \Big\} 
\end{align}
by further simplification we get
\begin{align}
-\partial_t \Psi(t,\vx) =& \min\limits_{\vu_t} \Big\{ L(t,\vx,\vu_t) + \nabla_{\vx}\Psi^T(t,\vx) \left( \vf(t,\vx)+\vG(t,\vx)\vu_t \right) \notag \\ 
& + \frac{\sigma}{2} \nabla_{\vx}\Psi^T(t,\vx) \vC(t,\vx_t)\Sigma \vC^T(t,\vx_t) \nabla_{\vx}\Psi(t,\vx) + \frac{1}{2} Tr [\nabla_{\vx\vx}\Psi(t,\vx) \vC(t,\vx_t) \Sigma \vC^T(t,\vx_t)] \Big\} 
\end{align}
If we assume that the cost function is quadratic with respect to control input
$\vu_t$ as
\begin{equation}
L(t,\vx,\vu_t) = \phi(t,\vx) + \frac{1}{2} \vu_t^T \vR(t) \vu_t + \vu_t^T \vr(t,x)
\end{equation}
the optimal control input will be
\begin{equation}
\vu^*(t,\vx) = - \vR(t)^{-1} \left( \vr(t,x) + \vG^T(t,\vx) \nabla_{\vx}\Psi(t,\vx) \right)
\end{equation}
and the HJB equation will be
\begin{equation}
-\partial_t \Psi = \phi -\frac{1}{2}\vr^T \vR^{-1} \vr + \nabla_{\vx}\Psi^T \left( \vf - \vG\vR^{-1}\vr \right) - \frac{1}{2} \nabla_{\vx}\Psi^T \left( GR^{-1}G^T -\sigma \vC\Sigma\vC^T \right) \nabla_{\vx}\Psi + \frac{1}{2} Tr [\nabla_{\vx\vx}\Psi \vC\Sigma\vC^T]
\end{equation}

%%%%%%%%%%%%%%%%%%%%%%%%%%%%%%%%%%%%%%%%%%%%%%%%%%%%%%%%%%%%%%%%%%%%%%%%%%%%%%%%
\section{Appendix C}
\textbf{\textit{Theorem 2}}: 
The solution to the optimal control problem defined in Equations\!
(\ref{eq:dynamics_linear_approximation}-\ref{eq:cost_quadratic_approximation_detail_2}) exists if $\left( \vB_t\vR_t^{-1}\vB_t^T -\sigma \vC_t\Sigma\vC_t^T \right)$ is positive semidefinite for all the \textit{t}s and the solution can be found as it follows
\begin{align}
-\dot{\vS}_t =& \vQ_{t} + \vA_t^T \vS_t + \vS_t^T \vA_t - \left( \vP_t^T + \vB_t^T \vS_t \right)^T \vR_t^{-1} \left( \vP_t^T + \vB_t^T \vS_t \right) \notag \\
 &+ \sigma \vS_t^T \vC_t\Sigma\vC_t^T \vS_t  \label{app_eq:riccati_Sm}\\
-\dot{\vs}_t =& \vq_{t} + \vA_t^T \vs_t - \left( \vP_t^T + \vB_t^T \vS_t \right)^T \vR_t^{-1} \left( \vr_t + \vB_t^T \vs_t \right) \notag \\ 
&+ \sigma  \vS_t^T \vC_t\Sigma\vC_t^T \vs_t \label{app_eq:riccati_Sv}\\
-\dot{s}_t =& q_{t} - \frac{1}{2} \left( \vr_t + \vB_t^T \vs_t \right)^T \vR_t^{-1} \left( \vr_t + \vB_t^T \vs_t \right)  + \frac{1}{2} Tr\left[ \vS(t) \vC_t\Sigma\vC_t^T \right] \notag \\
&+ \frac{\sigma}{2} \vs_t^T \vC_t\Sigma\vC_t^T \vs_t   \label{app_eq:riccati_s}
\end{align}
with the final values $\vS_{t_f} = \vQ_{f}$, $\vs_{t_f} = \vq_{f}$, and $s_{t_f}
= q_{f}$. The optimal control is
\begin{align}
&\delta\vu^*(t,\vx) = \vl(t) + \vL(t) \delta\vx \label{app_eq:optimal_control_update}\\
&\vl(t) = - \vR_t^{-1} \left( \vr_t + \vB_t^T \vs_t \right) \label{app_eq:optimal_control_l}\\
&\vL(t) = - \vR_t^{-1} \left( \vP_t^T + \vB_t^T \vS_t \right) \label{app_eq:optimal_control_L}
\end{align}

\textbf{\textit{Proof}}: 
The approximate optimal control problem defined in Equations\! (\ref{eq:dynamics_linear_approximation}-\ref{eq:cost_quadratic_approximation_detail_2}) can be solved by the use of Equation \eqref{app_eq:nonlinear_utility_pde}. We will make the following Ansatz for $\Psi(t,\delta\vx)$ to solve PDE
\begin{align}
&\Psi(t,\delta\vx) = s(t) + \vs(t)^T\delta\vx + \frac{1}{2}\delta\vx^T\vS(t)\delta\vx \\ 
&\partial_t \Psi(t,\delta\vx) = \dot{s}(t) + \dot{\vs}(t)^T\delta\vx + \frac{1}{2}\delta\vx^T\dot{\vS}(t)\delta\vx \\
&\nabla_{\vx}\Psi_{\delta x}(t,\delta\vx) = \vs(t) + \vS(t)\delta\vx  \\
&\nabla_{\vx\vx}\Psi_{\delta x  \delta x}(t,\delta\vx) = \vS(t) 
\end{align} 

Then we will have 
\begin{align}
&-\dot{s}_t - \dot{\vs}_t^T\delta\vx - \frac{1}{2}\delta\vx^T\dot{\vS}_t\delta\vx = 
q_{t} -\frac{1}{2} \vr_t^T \vR_t^{-1} \vr_t + \delta\vx^T \left(\vq_{t} - \vP_t \vR_t^{-1} \vr_t \right) + \frac{1}{2} \delta\vx^T \left( \vQ_{t}-\vP_t \vR_t^{-1} \vP_t^T \right) \delta\vx \notag \\
& + \left( \vs_t + \vS_t\delta\vx \right)^T \left( \vA_t\delta\vx - \vB_t \vR_t^{-1} \vr_t - \vB_t \vR_t^{-1} \vP_t^T \delta\vx \right) - \frac{1}{2} \left( \vs_t + \vS_t\delta\vx \right)^T \left( \vB_t\vR_t^{-1}\vB_t^T -\sigma \vC_t\Sigma\vC_t^T \right) \left( \vs_t + \vS_t \delta\vx \right)
+ \frac{1}{2} Tr\left[ \vS(t) \vC_t\Sigma\vC_t^T \right] 
\end{align}
we can rearrange the above equation as
\begin{align}
&-\dot{s}_t - \dot{\vs}_t^T\delta\vx - \frac{1}{2}\delta\vx^T\dot{\vS}_t\delta\vx = \notag \\
&q_{t} - \frac{1}{2} \vr_t^T \vR_t^{-1} \vr_t - \vs_t^T \vB_t \vR_t^{-1} \vr_t - \frac{1}{2} \vs_t^T \left( \vB_t\vR_t^{-1}\vB_t^T -\sigma \vC_t\Sigma\vC_t^T \right) \vs_t + \frac{1}{2} Tr\left[ \vS(t) \vC_t\Sigma\vC_t^T \right]  \notag \\
&+ \left(\vq_{t} - \vP_t \vR_t^{-1} \vr_t \right)^T \delta\vx + \vs_t^T \left( \vA_t - \vB_t \vR_t^{-1} \vP_t^T \right) \delta\vx - \vr_t^T \vR_t^{-1} \vB_t^T \vS_t \delta\vx - \vs_t^T \left( \vB_t\vR_t^{-1}\vB_t^T -\sigma \vC_t\Sigma\vC_t^T \right) \vS_t \delta\vx \notag \\
&+ \frac{1}{2} \delta\vx^T \left( \vQ_{t}-\vP_t \vR_t^{-1} \vP_t^T \right) \delta\vx + \delta\vx^T \vS_t^T \left( \vA_t - \vB_t \vR_t^{-1} \vP_t \right) \delta\vx - \frac{1}{2} \delta\vx^T \vS_t^T \left( \vB_t\vR_t^{-1}\vB_t^T -\sigma \vC_t\Sigma\vC_t^T \right) \vS_t \delta\vx 
\end{align}

By equating the coefficient of $\delta\vx$, we will have the following equations.
\begin{align}
-\dot{\vS}_t =& \vQ_{t} -\vP_t \vR_t^{-1} \vP_t^T + \left( \vA_t - \vB_t \vR_t^{-1} \vP_t^T \right)^T\vS_t + \vS_t^T \left( \vA_t - \vB_t \vR_t^{-1} \vP_t^T \right) - \vS_t^T \left( \vB_t\vR_t^{-1}\vB_t^T -\sigma \vC_t\Sigma\vC_t^T \right) \vS_t \\
-\dot{\vs}_t =& \vq_{t} - \vP_t \vR_t^{-1} \vr_t + \left( \vA_t - \vB_t \vR_t^{-1} \vP^T_t \right)^T \vs_t - \vS_t^T \vB_t \vR_t^{-1} \vr_t - \vS_t^T \left( \vB_t\vR_t^{-1}\vB_t^T -\sigma \vC_t\Sigma\vC_t^T \right) \vs_t  \\
-\dot{s}_t =& q_{t} - \frac{1}{2} \vr_t^T \vR_t^{-1} \vr_t - \vs_t^T \vB_t \vR_t^{-1} \vr_t - \frac{1}{2} \vs_t^T \left( \vB_t\vR_t^{-1}\vB_t^T -\sigma \vC_t\Sigma\vC_t^T \right) \vs_t + \frac{1}{2} Tr\left[ \vS(t) \vC_t\Sigma\vC_t^T \right]  
\end{align}
with final values
\begin{equation}
\vS_{t_f} = \vQ_{t_f}, \quad 
\vs_{t_f} = \vq_{t_f}, \quad
  s_{t_f} = q_{t_f}. 
\end{equation}
and the optimal control
\begin{equation}
\delta\vu^*(t,\vx) = - \vR_t^{-1} \left( \vr_t + \vB_t^T \vs_t \right)
- \vR_t^{-1} \left( \vP_t^T + \vB_t^T \vS_t \right) \delta\vx
\end{equation}

These equations will have solutions if $\left( \vB_t\vR_t^{-1}\vB_t^T -\sigma
\vC_t\Sigma\vC_t^T \right)$ is positive semidefinite for all \textit{t}. We
can further simplify these equations by regrouping them as
\begin{align}
-\dot{\vS}_t =& \vQ_{t} + \vA_t^T \vS_t + \vS_t^T \vA_t - \left( \vP_t^T + \vB^T \vS_t \right)^T \vR_t^{-1} \left( \vP_t^T + \vB^T \vS_t \right) + \sigma \vS_t^T \vC_t\Sigma\vC_t^T \vS_t \\
-\dot{\vs}_t =& \vq_{t} + \vA_t^T \vs_t - \left( \vP_t^T + \vB_t^T \vS_t \right)^T \vR_t^{-1} \left( \vr_t + \vB_t^T \vs_t \right) + \sigma  \vS_t^T \vC_t\Sigma\vC_t^T \vs_t  \\
-\dot{s}_t =& q_{t} - \frac{1}{2} \left( \vr_t + \vB_t^T \vs_t \right)^T \vR_t^{-1} \left( \vr_t + \vB_t^T \vs_t \right) + \frac{1}{2} Tr\left[ \vS(t) \vC_t\Sigma\vC_t^T \right] + \frac{\sigma}{2} \vs_t^T \vC_t\Sigma\vC_t^T \vs_t  
\end{align}

\end{document}